\documentclass[letter,apj]{emulateapj-rtx4}
\pdfoutput=1
\usepackage{graphicx}
\usepackage{enumerate}
\usepackage{amssymb, amsmath}
\usepackage{natbib}
\usepackage{color}
\usepackage{ulem}

\newcommand{\ias}{1}
\newcommand{\princeton}{2}
\newcommand{\sagan}{3}

\begin{document}

\title{The Age and Age Spread of the Praesepe and Hyades Clusters: a Consistent, $\sim$800 Myr Picture from Rotating Stellar Models}
\author{Timothy D.~Brandt\altaffilmark{\ias, \sagan} \&
Chelsea X.~Huang\altaffilmark{\princeton}
}

\altaffiltext{\ias}{School of Natural Sciences, Institute for Advanced Study, Princeton, NJ, USA.}
\altaffiltext{\princeton}{Department of Astrophysical Sciences, Princeton University, Princeton, NJ, USA.}
\altaffiltext{\sagan}{NASA Sagan Fellow}

\begin{abstract}
We 
fit the upper main sequence of the Praesepe and Hyades open clusters using stellar models with and without rotation.  When neglecting rotation, we find that no single isochrone can fit the entire upper main sequence at the clusters' spectroscopic metallicity: more massive stars appear, at high significance, to be younger than less massive stars.  This discrepancy is consistent with earlier studies, but vanishes when including stellar rotation.  The entire upper main sequence of both clusters is very well-fit by a distribution of 800 Myr-old stars with the spectroscopically measured $[{\rm Fe/H}] = 0.12$.  The increase over the consensus age of $\sim$600-650 Myr is due both to the revised Solar metallicity (from $Z_\odot \approx 0.02$ to $Z_\odot \approx 0.014$) and to the lengthening of main sequence lifetimes and increase in luminosities with rapid rotation.  
Our results show that rotation can remove the need for large age spreads in intermediate age clusters, and that these clusters may be significantly older than is commonly accepted.  A Hyades/Praesepe age of $\sim$800 Myr would also require a recalibration of rotation/activity age indicators.

\end{abstract}

\section{Introduction} \label{sec:intro}

Stellar models have long been fit to star clusters to determine their ages and metallicities; fine grids of evolutionary models are now available from several groups \citep{Yi+Demarque+Kim+etal_2001, Girardi+Bertelli+Bressan+etal_2002, Pietrinferni+Cassisi+Salaris+etal_2004, Dotter+Chaboyer+Jevremovic+etal_2008}.  For young clusters, techniques including fitting models of pre-main sequence contraction \citep{Siess+Dufour+Forestini_2000, Pecaut+Mamajek+Bubar_2012} and lithium depletion \citep{Jeffries+Naylor+Mayne+etal_2013, Binks+Jeffries_2014} can provide reliable ages.  These techniques are not applicable to intermediate age clusters, from several hundred Myr to several Gyr.  Models of stellar evolution are often the best dating technique available in these cases \citep{Soderblom_2010}.  The ages from stellar modeling are then used to calibrate secondary age indicators, including those based on stellar rotation and activity \citep{Noyes+Hartmann+Baliunas+etal_1984, Barnes_2007, Mamajek+Hillenbrand_2008, Epstein+Pinsonneault_2014}.

The Hyades and Praesepe (the Beehive cluster) are two of the best-studied nearby open clusters.  As such, they currently serve as benchmarks for stellar ages and stellar modeling \citep{Soderblom_2010}.  The clusters are both part of the Hyades supercluster, and seem to share an age, as determined both from isochrone fitting \citep{Perryman+Brown+Lebreton+etal_1998, Salaris+Weiss+Percival_2004, Fossati+Bagnulo+Landstreet+etal_2008} and from stellar rotation \citep[gyrochronology,][]{Douglas+Agueros+Covey+etal_2014, Kovacs+Hartman+Bakos+etal_2014}.  The Hyades and Praespe also have very similar (possibly identical) chemical compositions \citep{Boesgaard+Roper+Lum_2013, Taylor+Joner_2005}, hinting at formation in a single molecular cloud or cloud complex.  

The age, and even the coevality, of the Hyades and Praesepe are not completely settled.  Stellar isochrones seem to suggest an age range of several hundred Myr \citep{Eggen_1998}, with the main-sequence turnoff giving an age of $\sim$600-650 Myr for the most massive members \citep{Perryman+Brown+Lebreton+etal_1998, Fossati+Bagnulo+Landstreet+etal_2008}.  Other dating methods have been applied to the Hyades, including white dwarf cooling tracks \citep{DeGennaro+vonHippel+Jefferys+etal_2009}, giving ages consistent with those from the upper main sequence.  Ultimately, however, these alternatives rely on the same stellar models as ordinary isochrone fitting.

On a different ground, various authors noticed that the presence of extended main sequence turnoffs in 1--2 Gyr-old clusters, indicating an intracluster age spread of several hundred Myr \citep{Mackey+Nielsen_2007,Milone+Bedin_etal_2009, Li+deGrijs+Deng_2014}. The effect of stellar rotation on the color-magnitude diagrams has been investigated as a possible solution to resolve this problem.   \citet{Bastian+de_Mink_2009} pointed out that the effects of rapid rotation on stellar evolutionary tracks could modify the isochrone, leading to a redder and cooler main sequence turnoff.  However, \citeauthor{Bastian+de_Mink_2009} neglected the extension of the main sequence lifetime due to rotation. \citet{Girardi+Eggenberger+Miglio_2011} pointed out that this change in the stellar lifetime is important, and used the Geneva stellar evolution code \citep{Eggenberger+Meynet_etal_2008} to get hotter and bluer 1.58 Gyr isochrones with a uniform rotation rate.  These isochrones, like those with core convective overshoot but without rotation, did not show a significant extended main sequence turnoff.  \citet{Yang+Meng+Liu_2013}, however, computed their own evolutionary tracks including rotation, and did find extended main sequence turn offs in younger clusters. 

While we do not intend to resolve the controversy over the extended main sequence problem in this article, those efforts hint that rotation might be a key to resolving the age spread observed by \citet{Eggen_1998} (Figure 1) in the Hyades.  Rotation has long been known to have a potentially large effect on the evolutionary tracks of stars $\gtrsim$1.5 $M_\odot$ \citep{Meynet+Maeder_2000}.  This is distinct from the B\"ohm-Vetense gap \citep{Bohm-Vitense_1970} seen in the Hyades by \cite{deBruijne+Hoogerwerf+deZeeuw_2000}, which occurs at the onset of surface convection (and of efficient magnetic braking).  

Recently, \citet{Brandt+Huang_2015}, henceforth BH15, applied new rotating stellar models \citep{Ekstrom+Georgy+Eggenberger+etal_2012, Georgy+Ekstrom+Eggenberger+etal_2013} to several clusters, including the Hyades. They found a best-fit Hyades age of $\sim$800 Myr from the upper main-sequence turnoff, significantly older than the current consensus.  We apply the same methodology as BH15 to Praesepe as a consistency check. We then further investigate whether rotation can resolve both the apparent spread in Hyades ages and the inconsistency of our older Hyades age with the consensus age in the literature.  

We organize the paper as follows.  In Section \ref{sec:method} we review our methodology from BH15, referring to that paper for a more thorough description.  In Section \ref{sec:consistency}, we present a statistical derivation of the criterion we use to assess the consistency of the isochrone-based ages between individual cluster members.  We discuss the selection of our cluster samples in Section \ref{sec:sample}.  Section \ref{sec:results} contains our results; we conclude with Section \ref{sec:conclusions}.

\section{Methodology} \label{sec:method}

We apply the Bayesian method used in BH15 to derive the ages of the Hyades and Praesepe open clusters.  We summarize the method here, and refer the reader to that paper for a more thorough description.  Our method takes as its input {\it Tycho}-2 $B_T V_T$ magnitudes \citep{Hog+Fabricius+Makarov+etal_2000}, parallax, rotational $v \sin i$, and a prior probability distribution of metallicity, and returns posterior probability distributions of mass, metallicity, age, and inclination.  In the rest of this paper, we will concern ourselves only with the posterior probability distributions of age and metallicity.  We marginalize over the mass and inclination distributions star-by-star.

We use the new rotating isochrones of \cite{Ekstrom+Georgy+Eggenberger+etal_2012} and \cite{Georgy+Ekstrom+Eggenberger+etal_2013} as our stellar models.  These are available at a grid of rotation rates ranging from 0 to 95\% of breakup, stellar masses from 1.7 to 15 $M_\odot$, and metallicities from $Z=0.002$ to $Z=0.014$ ($[{\rm Fe/H}] = -0.85$ to $[{\rm Fe/H}] = 0$).  The mass grid is extremely coarse, so we interpolate these models to higher resolution using nonrotating stellar isochrones.  We use the PARSEC models \citep{Girardi+Bertelli+Bressan+etal_2002} for this purpose.  We compute the corrections induced by rotation as a function of mass, metallicity, and rotation rate, and interpolate (and even extrapolate) these coefficients onto the much finer PARSEC grid.  The validity of this step relies on the fact that the rotational correction term is a very weak function of the other stellar parameters.  

The rotating stellar models adopt a core overshooting parameter, the ratio of the convective overshoot to the pressure scale height, of 0.1.  This value was chosen to reproduce the observed width of the main sequence when including rotation \citep{Ekstrom+Georgy+Eggenberger+etal_2012}.  Core overshooting extends the effective size of the convective core, allowing more of the star to be burned.  Rotation achieves a similar effect, but by mixing hydrogen into the core rather than by simply making the core larger \citep{Talon+Zahn+Maeder+etal_1997}.   \cite{Schaller+Schaerer+Meynet+etal_1992} found that an overshoot parameter of 0.2 was needed to reproduce the observed main sequence width without rotation.  This degree of overshooting is in mild tension with asteroseismology of two slowly rotating B stars \citep{Papics+Moravveji+Aerts+etal_2014, Moravveji+Aerts+Papics+etal_2015}.  The recent asteroseismic results favor an overshooting parameter between 0.1 and 0.2, though they find better agreement with exponentially decreasing core overshoot than with a step function as commonly used in grids of stellar models (including the models we use here).

The \cite{Georgy+Ekstrom+Eggenberger+etal_2013} stellar models give the stellar equatorial radius, luminosity, and oblateness.  We assume that the star may described by a Roche model and use \cite{Lara+Rieutord_2011} to compute the local effective temperature and gravity everywhere on the stellar surface.  We then use the full specific intensities of the \cite{Castelli+Kurucz_2004} model atmospheres and the {\it Tycho}-2 bandpasses as determined by \cite{Bessell+Murphy_2012} to compute synthetic $B_T V_T$ photometry as a function of orientation.  The Solar abundances have been updated since the \cite{Castelli+Kurucz_2004} models were published, with \cite{Asplund+Grevesse+Sauval+etal_2009} finding a Solar metal abundance nearly 0.2 dex lower than \cite{Anders+Grevesse_1989}, and a Solar iron abundance 0.17 dex lower.  Because the \cite{Georgy+Ekstrom+Eggenberger+etal_2013} stellar models use the new abundances, we adopt the ATLAS9 atmospheres with $[{\rm Fe/H}] = -0.1$ relative to the \cite{Anders+Grevesse_1989} composition for a $Z_\odot$ star.  This provides an approximate match between the star's bulk and photospheric compositions.  It could, however, introduce systematic differences in metallicity at a level $\Delta [{\rm Fe/H}] \sim 0.1$ relative to earlier results, and should be kept in mind when comparing different authors' stellar tracks.  At an age of $\sim$700 Myr, a decrease of 0.1 dex in $[{\rm Fe/H}]$ can, in some ways, mimic an increase of up to $\sim$100 Myr in age.

Finally, we fit the synthetic photometry and predicted $v \sin i$ to the observed photometry, rotation, and parallax.  We adopt the appropriate priors: masses drawn from a \cite{Salpeter_1955} initial mass function, random orientations, and a uniform prior in volume.  We use a prior in stellar rotation that closely matches the observed distribution in young, massive field stars \citep{Zorec+Royer_2012}.  We add a systematic error of 30 km\,s$^{-1}$ in $v \sin i$ and 0.005 mag in $B_T$ and $V_T$ to account for errors induced by our interpolations and finite grid spacing.  

When computing the age and metallicity posterior probability distribution of a cluster, we multiply the posterior probability distributions for the individual stars.  This implicitly assumes the cluster to have a single age and composition, i.e., a common origin.  We marginalize over the other parameters ($M$, $v \sin i$, and inclination) separately for each star.  In the next section, we derive a consistency criterion, equivalent to the usual $\chi^2$ test, to assess whether the posterior probability distributions for the individual stars really are consistent with a common origin.

\section{Testing the Consistency of Posterior Probability Distributions} \label{sec:consistency}

When determining the posterior probability distribution in age and composition for a cluster, we multiply the probability distributions for the individual stars.  Each of these distributions is computed using the Bayesian formalism detailed in BH15 and summarized in Section \ref{sec:method}, and marginalized over mass, rotation, and inclination.  By multiplying them together, we implicitly assume that the distributions are independent in at least one parameter and that they are consistent with a single cluster age and metallicity.  Our Bayesian formalism, however, does not provide a test of these assumptions.

The assumptions of independence and consistency are the same as those used in an ordinary $\chi^2$ analysis.  This fact allows us to use the $\chi^2$ test as a check on the consistency of the probability distributions and, by extension, on the ability of the stellar models to reproduce the entire cluster at a single age and composition.  We now derive an approximate relationship between the product of the peaks of the individual distribution functions and the peak of their product under the assumptions of independence and consistency.  Unsurprisingly, this ratio turns out to be proportional to the likelihood function.  The $\chi^2$ test has no effect on our posterior probability distributions, but only provides a consistency check on the stellar models themselves.

We begin with the two-dimensional age-metallicity probability distributions for each star, the outputs from the Bayesian analysis of Section \ref{sec:method} and BH15.  We make two approximations, well-satisfied in practice, to reduce these distributions to one dimension:
\begin{enumerate}
\item The covariance between age $\tau$ and metallicity $[{\rm Fe/H}]$ is the same for all stars; and
\item The posterior probability distribution on $[{\rm Fe/H}]$ is entirely determined by our prior.
\end{enumerate}
In this limit, the constraints provided by the stellar models are one-dimensional Gaussians extending parallel to a single line in $\tau$-$[{\rm Fe/H}]$ space (the covariance matrix has one eigenvalue much larger than the other).  We define the variable $x$ to be the linear combination of $\tau$ and $[{\rm Fe/H}]$ that we are actually constraining, the variable that runs in the direction of minimum covariance.  We obtain the total posterior probability distribution function by multiplying these one-dimensional Gaussians, and then multiplying by the metallicity prior.  

In our case we have $N_{\rm stars}$ estimates $x_i$ of the actual $x$, where $x$ is a linear combination of age and metallicity (an eigenvector of the covariance matrix).  For $N$ Gaussian measurements of a value $x_0$, the likelihood ${\cal L}$ of $x$ is 
\begin{align}
{\cal L} \propto \prod_i \exp \left[ -\frac{(x_i - x_0)^2}{2\sigma_i^2} \right] 
= \exp \left[ - \sum_i \frac{(x_i - x_0)^2}{2\sigma_i^2} \right]~,
\label{eq:chi2dist}
\end{align}
i.e., the exponential of one-half the $\chi^2$ distribution.  Because the true value $x_0$ is not known, the relevant $\chi^2$ distribution has $N_{\rm stars} - 1$ degrees of freedom.  We can then use the standard $\chi^2$ test to check the consistency of the individual posterior probability distributions with one another.

\section{Sample Selection} \label{sec:sample}

We select our Hyades stars from the analysis of \cite{Perryman+Brown+Lebreton+etal_1998}, rejecting the blue straggler HIP 20648 and known spectroscopic binaries, and keeping stars with a measured $B_T - V_T < 0.3$ in the {\it Tycho}-2 catalog \citep{Hog+Fabricius+Makarov+etal_2000}.  \cite{Perryman+Brown+Lebreton+etal_1998} rejected several other stars that lie above the main sequence.  We reject these same stars (HIP 20711, HIP 20901, HIP 21670, and HIP 22565) to enable a direct comparison with earlier results.  Including all of these targets (apart from the blue straggler HIP 20648) would have almost no effect on our results.  We obtain rotational velocities from the catalog compiled by \cite{Glebocki+Gnacinski_2005}, and adopt the {\it Hipparcos} parallaxes and errors for all targets \citep{vanLeeuwen_2007}.  Our final Hyades sample includes 14 stars.

We select our Praesepe sample from the candidates listed in \cite{Wang+Chen+Lin+etal_2014} with $B_T-V_T < 0.3$ (without correcting for extinction) and without indications of a close binary companion.  After correcting for $E(B-V) = 0.027$ mag \citep{Taylor_2006}, our Praesepe sample has a slightly bluer cutoff than our Hyades sample.  We exclude two stars from this intial sample: the blue straggler 40 Cnc, and the otherwise unremarkable $\delta$ Scuti variable BS Cnc, which lies significantly below the cluster main sequence.  This leaves a Praesepe sample of 24 stars.  We adopt a distance to the cluster of $179 \pm 3$ pc \citep{Gaspar+Rieke+Su+etal_2009}, which we use for each of the individual stars.  This is slightly higher than the reported uncertainty of $\pm 2$ pc (1$\sigma$) to allow for the small, but finite, extent of the cluster core.  Most of these stars have $v \sin i$ measurements in the catalog of \cite{Glebocki+Gnacinski_2005}.

For this analysis, we adopt a metallicity prior of $[{\rm Fe/H}] = 0.12 \pm 0.04$ for both clusters.  This value has been measured for a large sample of Praesepe dwarfs \citep{Boesgaard+Roper+Lum_2013} and is fully consistent with the best measurements of the Hyades \citep{Taylor+Joner_2005}.  It is slighly higher than the value of $[{\rm Fe/H}] = 0.10 \pm 0.05$ adopted by BH15, but is well within the uncertainties.

\section{Results} \label{sec:results}

\subsection{The Age of Praesepe} \label{subsec:praesepe}

\begin{figure}
\includegraphics[width=\linewidth]{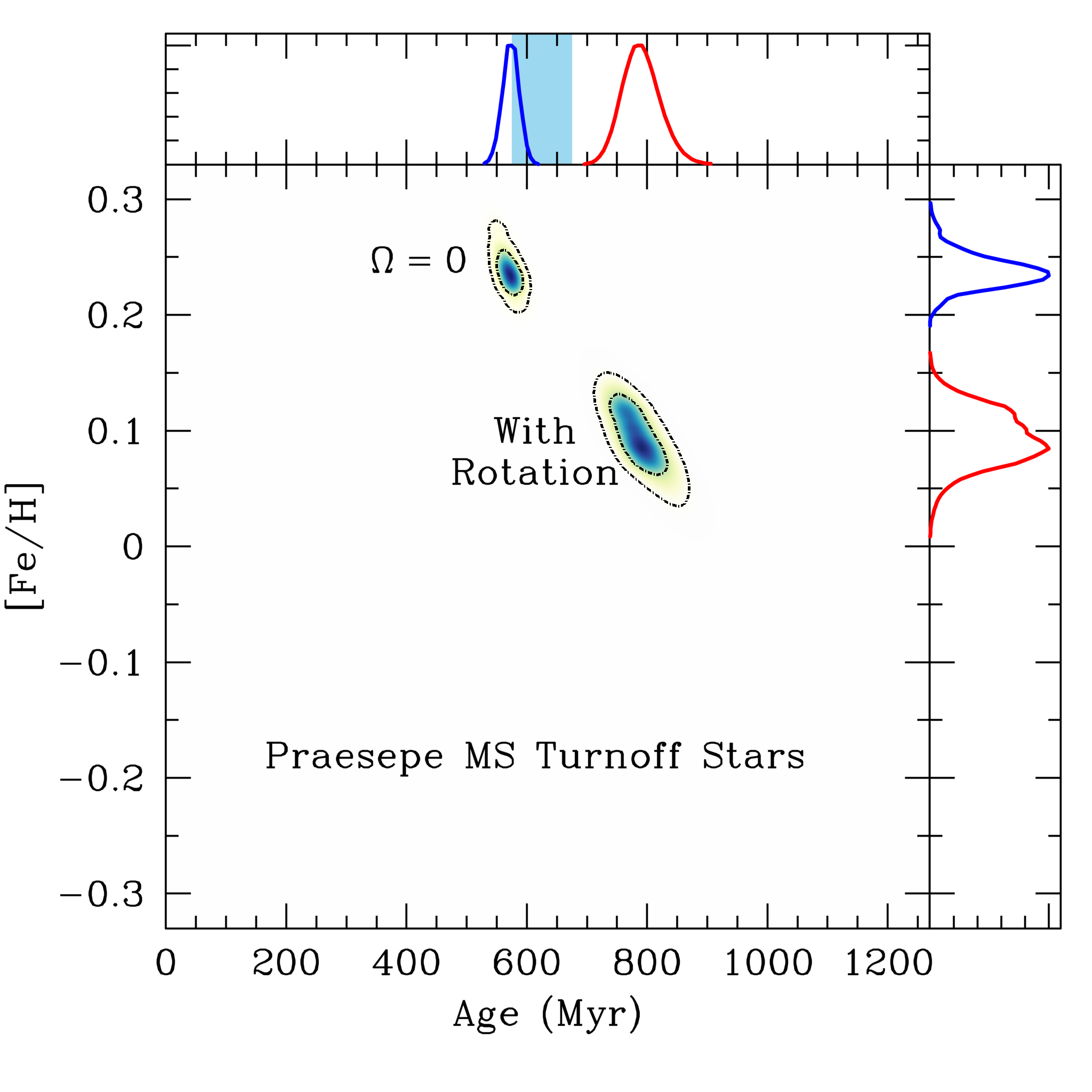}
\caption{The age of Praesepe computed by applying the method described in BH15 and summarized in Section \ref{sec:method} to 24 early-type likely members.  We have adopted a metallicity prior of $[{\rm Fe/H}] = 0.12 \pm 0.04$ \citep{Boesgaard+Roper+Lum_2013} and a reddening $E(B-V)=0.027$ mag \citep{Taylor_2006}.  Stellar models including rotation indicate an age of $810 \pm 70$ Myr, much older than the consensus age of $\sim$600 Myr.  Much of the increase is due to the lengthening of the main-sequence lifetime with rotation.  Nonrotating models cannot produce a single consistent age, finding a best fit isochrone at a metallicity at the extreme end of our prior.  }
\label{fig:praesepe}
\end{figure}

Figure \ref{fig:praesepe} shows the results of our analysis for 24 likely Praesepe members in the age-metallicity plane.  The dot-dashed contours enclose 68\% and 95\% of the posterior probability, respectively.  The metallicity posterior probability distribution is almost entirely determined by our prior of $[{\rm Fe/H}] = 0.12 \pm 0.04$.  Our marginalized posterior probability distribution on age, the red curve at the top of Figure \ref{fig:praesepe}, gives an age of $790 \pm 60$ Myr (2$\sigma$ limits).  This is consistent with the Hyades age given in BH15.  

As for the Hyades, and as discussed in BH15, this old age arises from a combination of the increase in main sequence of lifetime with rotation, as rotational mixing supplies the core with additional unburnt fuel \citep{Ekstrom+Georgy+Eggenberger+etal_2012}.  The blue curve, and lower-left color patch, show the posterior probability distribution with the age artificially scaled back by this factor.  A small additional effect arises from the increase in luminosity in the latter part of the main sequence.  Increasing a star's luminosity with rotation means that a rotating model overlaps a slightly more massive (and shorter-lived) nonrotating model in color-magnitude space.

The best-fit age when fixing the rotation rate in the stellar models to zero is much lower: $570 \pm 30$ Myr.  This value masks a larger problem with the nonrotating models: {\it the fitted ages of the more luminous stars are inconsistent with those of the less luminous stars}, a problem that the next subsection will discuss in detail.  The only way to achieve some measure of consistency is to use a metallicity 2-3$\sigma$ away from the mean of our prior.  According to the $\chi^2$ test, the nonrotating distributions for the individual stars are inconsistent with one another at 99.99\% confidence, while the rotating distributions disagree at 74\% significance, i.e., slightly more than the median disagreement expected purely by chance.  In other words, a single nonrotating isochrone cannot provide an adequate fit to Praesepe.  A rotating isochrone with $\Omega/\Omega_{\rm crit}$ approximating the observed distribution, on the other hand, provides an excellent fit.

\subsection{Hyades and Praesepe Ages by Stellar Luminosity} \label{subsec:consistency}

\begin{figure}
\includegraphics[width=\linewidth]{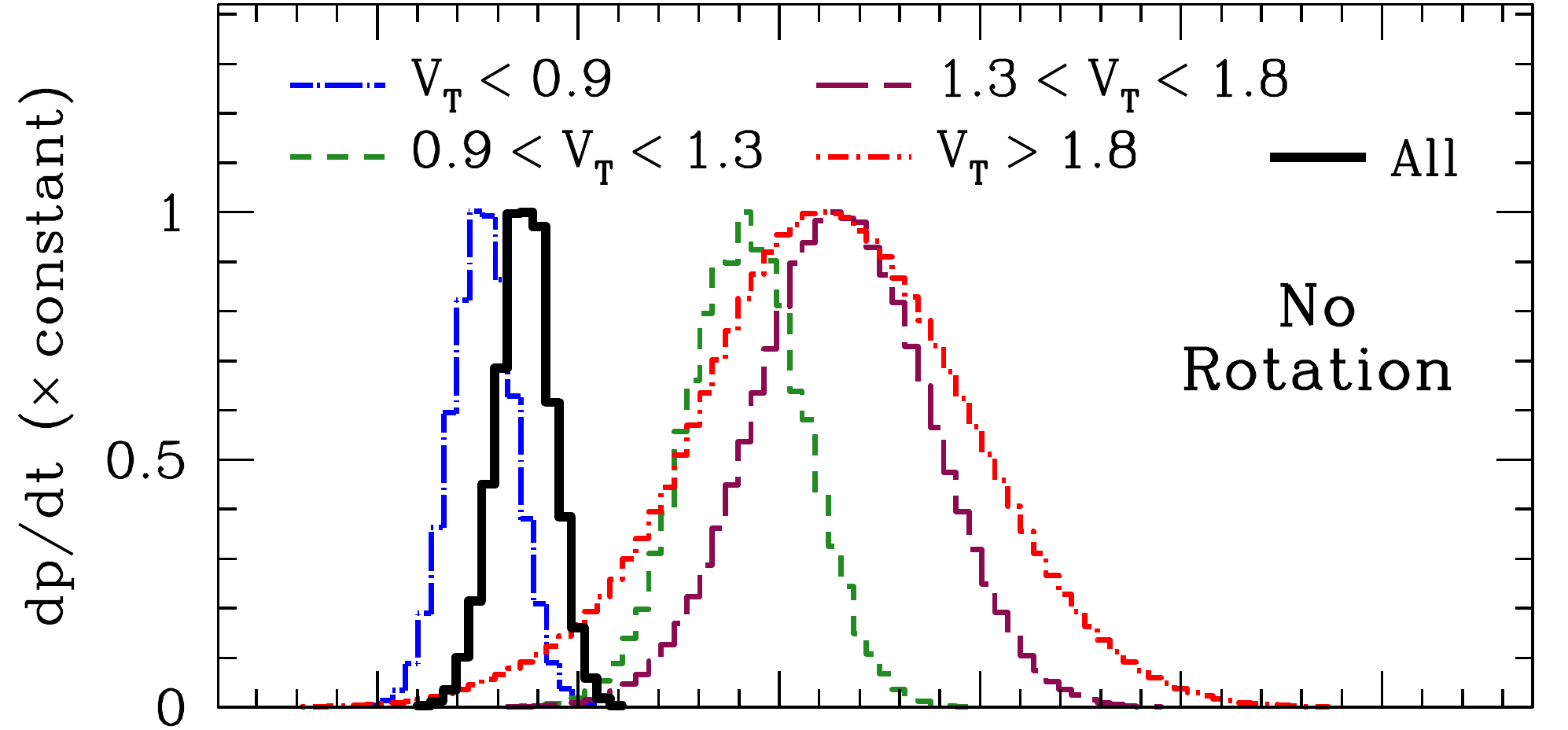}
\includegraphics[width=\linewidth]{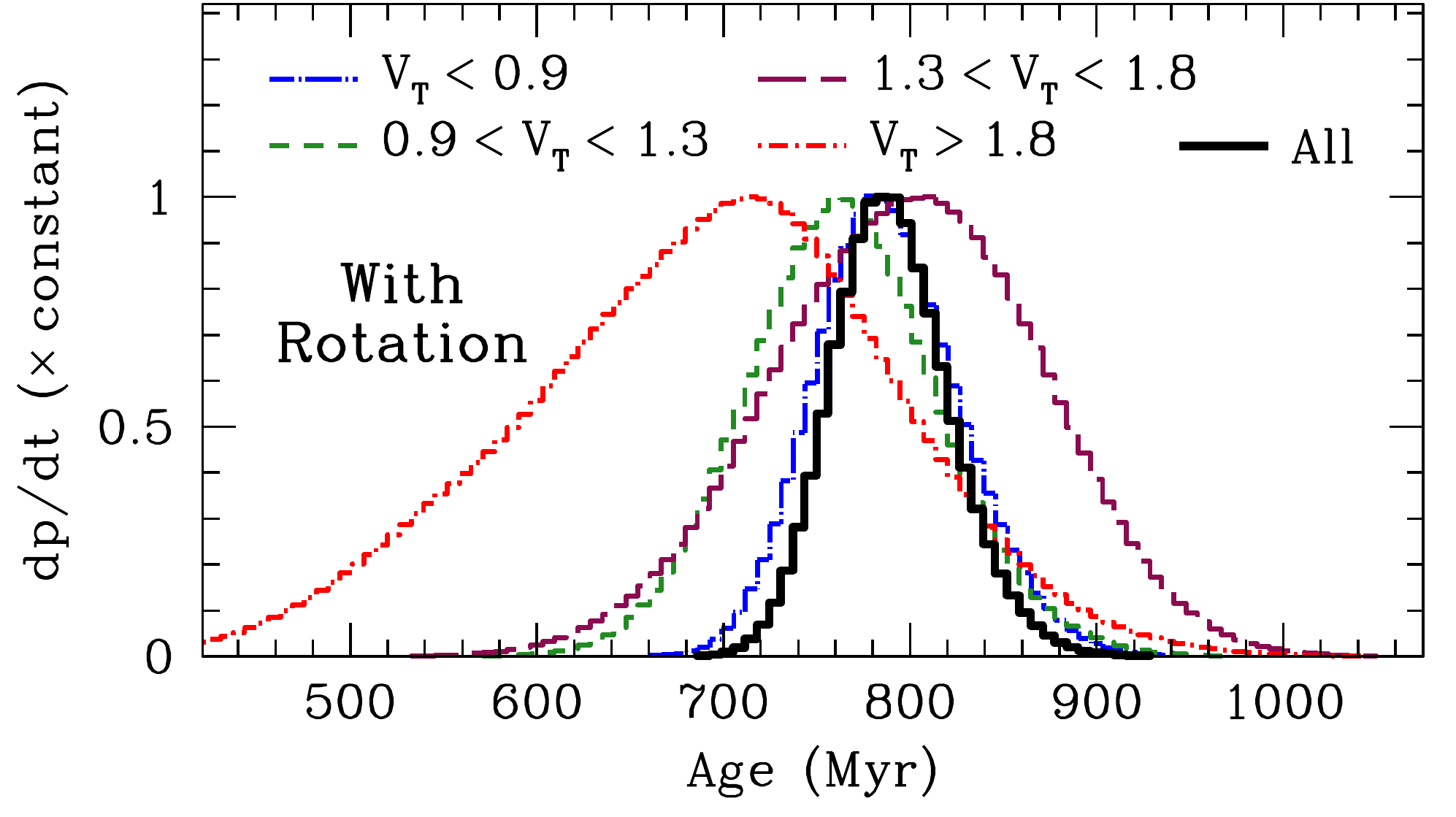}
\caption{Praesepe posterior probability distributions for subsets of the cluster, separated by luminosity.  The absolute magnitudes assume $E(B-V)=0.027$ mag and $d=179$ pc. Without rotation, the inferred age of the cluster depends on the part of the turnoff region is used (top panel).  The entire turnoff region is not consistent with a single age.  This inconsistency vanishes when accounting for rotation (bottom panel).  The entire cluster is consistent with an age of $790 \pm 60$ Myr (2$\sigma$), much older than the consensus age.}
\label{fig:praesepe_bymag}
\end{figure}

With nonrotating stellar models, our analysis for Praesepe fails the $\chi^2$ test (Section \ref{sec:consistency}): the cluster is not consistent with a single isochrone.  We now address this problem by separating our sample into four bins by stellar luminosity (corrected for $A_V = 0.084$).  The bin are chosen to have comparable statistical power and to be consistent with the Hyades bins we use below.  The most luminous bin, $V_T < 0.9$, has three stars, the next bin has four, the next six, and the least luminous bin has eleven stars, for a total of 24.  

The top panel of Figure \ref{fig:praesepe_bymag} clearly shows that the four bins of stars are inconsistent with being coeval.  The more massive, more luminous members require an age of $\sim$550 Myr (blue curve, close to the canonical age of the cluster), while the less luminous members require ages of 600-900 Myr.  The product of the distribution functions produces the thick black curve centered near 570 Myr.  Simply showing this curve masks the strong inconsistency between the component distributions, which, according to the $\chi^2$ test, have less than a 0.01\% chance of being so strongly inconsistent by chance.  

The bottom panel of Figure \ref{fig:praesepe_bymag} shows that the strong discrepancy between the age of the most and least luminous turnoff stars vanishes when accounting for stellar rotation.  All stars are consistent with a single age and composition.  The composition is perfectly consistent with the spectroscopic value of $[{\rm Fe/H}] = 0.12 \pm 0.04$ \citep{Boesgaard+Roper+Lum_2013}, but the age, $\sim$800 Myr, is much older than the canonical cluster age.  This age could be somewhat younger when accounting for systematic differences in the value of the Solar metallicity; at these ages, $\Delta [{\rm Fe/H}] = 0.1$ can roughly mimic a $\sim$100 Myr difference in age.  Eliminating core convective overshoot (currently $\alpha_{\rm OV}=0.1$) from the \cite{Georgy+Ekstrom+Eggenberger+etal_2013} could also produce a slightly younger age.  Asteroseismology of a slowly rotating B star, however, requires at least this modest degree of overshooting \citep{Moravveji+Aerts+Papics+etal_2015}.

\begin{figure}
\includegraphics[width=\linewidth]{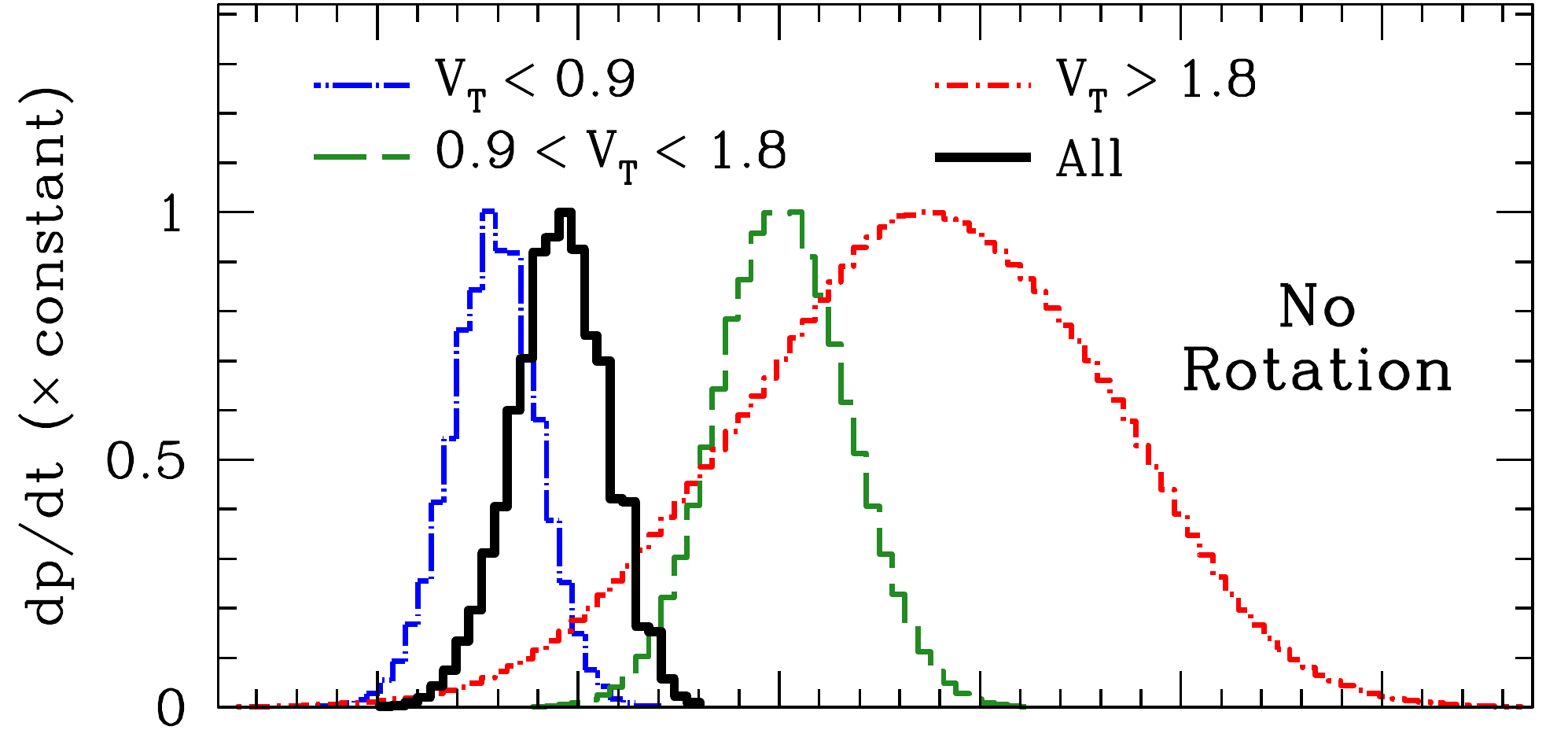}
\includegraphics[width=\linewidth]{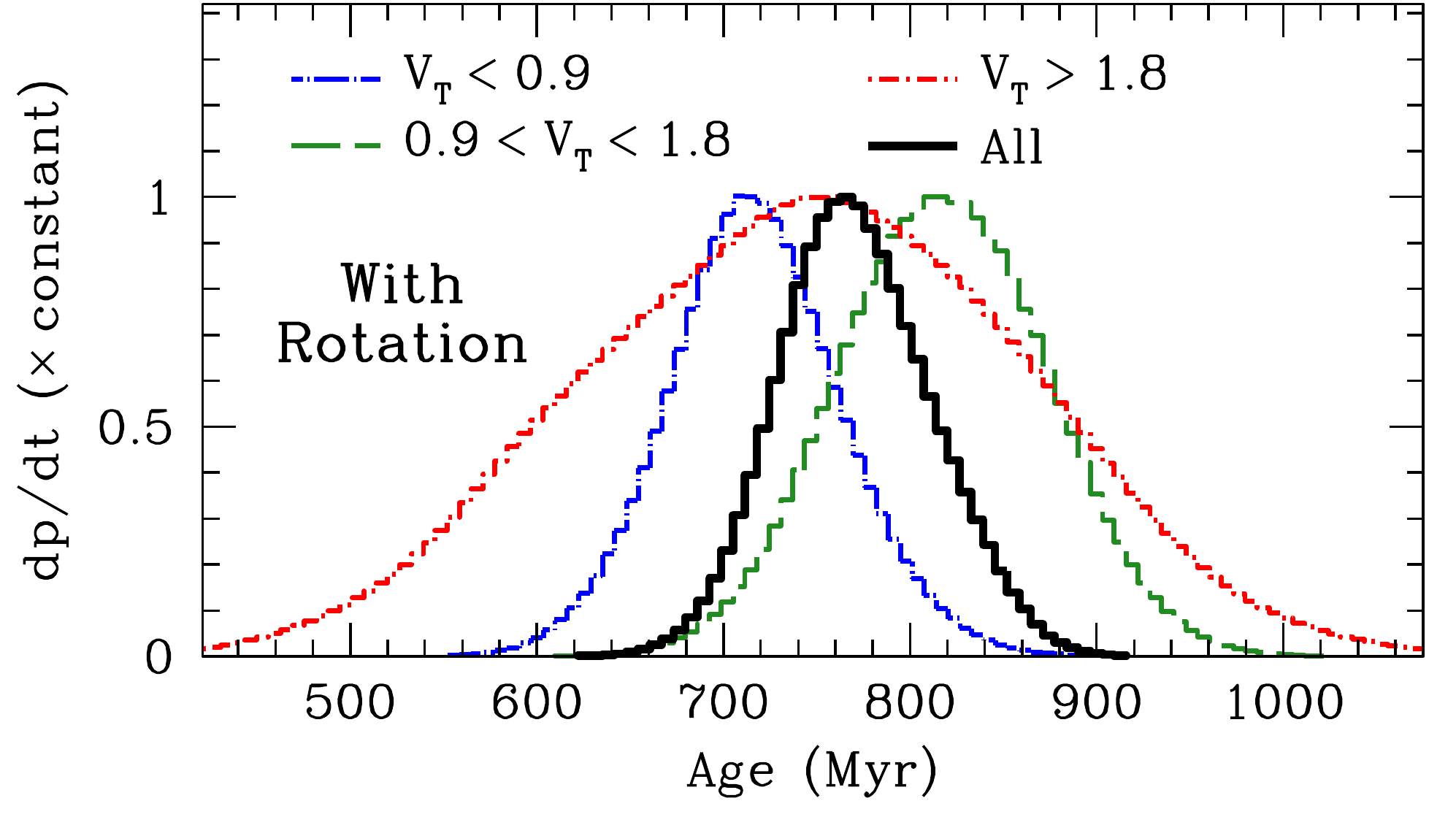}
\caption{Hyades posterior probability distributions for subsets of the cluster, separated by luminosity, and only using those stars that were also fit by \cite{Perryman+Brown+Lebreton+etal_1998}.  The top panel shows the results without rotation.  As for Praesepe, the best-fit age depends on the stellar luminosity, and the cluster is inconsistent with a single age.  Also as for Praesepe, the inconsistency vanishes when accounting for rotation.}
\label{fig:hyades_bymag}
\end{figure}

Figure \ref{fig:hyades_bymag} tells the same story as Figure \ref{fig:praesepe_bymag}, but for the Hyades rather than Praesepe.  Because the Hyades is a sparser cluster, we use only three bins in absolute $V_T$ magnitude.  The most luminous bin, $V_T<0.9$, has two stars, the middle bin has six, and the least luminous bin has seven, for a total of 15 stars.  These stars extend to slightly redder colors than the Praesepe sample (after correcting for Praesepe's extinction).

As for Praesepe, the Hyades stars are inconsistent with a single nonrotating isochrone, though not quite so strongly inconsistent.  The $\chi^2$ test gives a 0.4\% probability of at least the observed discrepancy occurring by chance.  The stars are only consistent with a single isochrone at high metallicity, however, $[{\rm Fe/H}] \approx 0.25$.  This is incompatible with the spectroscopic Praesepe and Hyades values and the revised Solar composition.  Constraining $[{\rm Fe/H}] = 0.12$ exactly, a $\chi^2$ test indicates that the stars are fully consistent with a single age when including rotation, and inconsistent at more than 99.99\% probability when excluding rotation.

\subsection{A Consistent, Coeval Picture with Rotation} \label{subsec:rotation}

\begin{figure}
\includegraphics[width=1.05\linewidth]{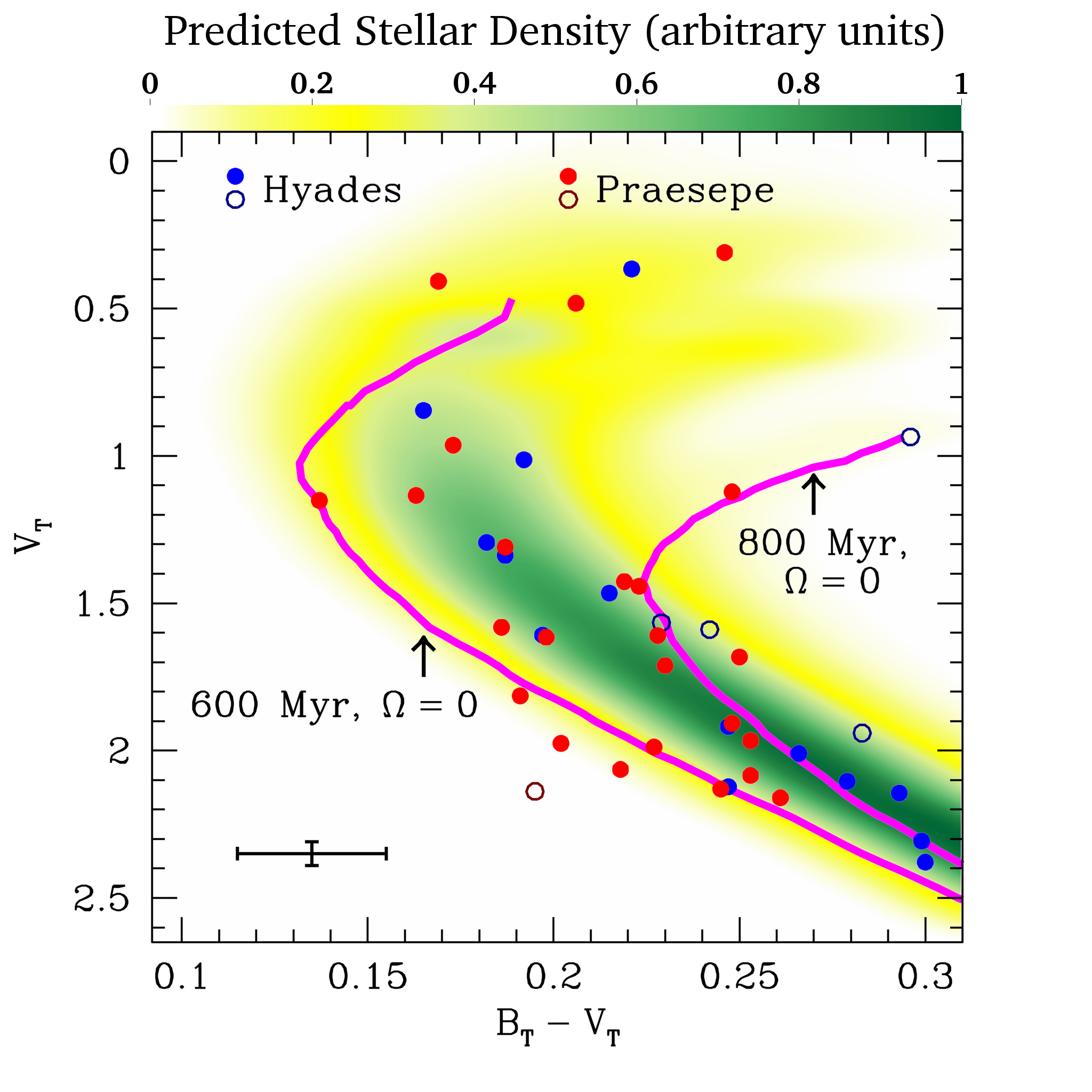}
\caption{Color-magnitude diagram of the Hyades (blue) and Praesepe (red) stars used in our analysis; cluster members excluded from the analysis in Sections \ref{subsec:praesepe} and \ref{subsec:consistency} are shown by open circles.  We have excluded the same Hyades members as \cite{Perryman+Brown+Lebreton+etal_1998} to enable a direct comparison.  Neither cluster is consistent with a single nonrotating isochrone, as indicated by the magenta curves representing 600 Myr and 800 Myr stellar populations with $[{\rm Fe/H}] = 0.12$.  A rotating 800 Myr isochrone at $[{\rm Fe/H}] = 0.12$ and with the observed distribution of rotation rates, convolved with the typical uncertainty (lower-left cross) and shaded by its predicted stellar density in the color-magnitude diagram, provides an excellent fit to all of the stars. }
\label{fig:HR}
\end{figure}

More than 15 years ago, \cite{Eggen_1998} pointed out that the ages of Hyades and Praesepe stars seemed to vary systematically with stellar mass.  This may also be seen in Figure 22 of \cite{Perryman+Brown+Lebreton+etal_1998}: the stars less luminous than $M_V \approx 1.8$ fit an older isochrone than the more massive turnoff stars.  This remained true even after \citeauthor{Perryman+Brown+Lebreton+etal_1998} rejected single stars lying above the main sequence.  The same disagreement between the age inferred from the most luminous turnoff stars and the rest of the upper main sequence may be seen in the lower-left panel of Figure 17 of \cite{David+Hillenbrand_2015}.  The most common solution is to accept the turnoff age and neglect the modest discrepancy at lower luminosities.  

Stellar rotation, however, enables a fit of the entire upper main sequence and turnoff.  The previous section showed this with stars divided into luminosity bins.  Figure \ref{fig:HR} shows the full color-magnitude diagram, with two nonrotating isochrones (600 and 800 Myr, with $[{\rm Fe/H}] = 0.12$) and one 800 Myr, $[{\rm Fe/H}] = 0.12$ rotating isochrone (yellow and green density plot).  The density plot has been convolved with the typical observational uncertainties, which are indicated by a cross in the lower-left of the image.  The units of the density plot are stars per unit $V_T$ per unit $B_T - V_T$.  The selection of candidate members is described in Section \ref{sec:sample}; open circles were excluded from the preceding section's analysis.  While increasing the metallicity improves the agreement without rotation, we find a best-fit $[{\rm Fe/H}] \approx 0.12$ with the rotating models, an essentially perfect match to the spectroscopic metallicity.

A rotating isochrone at a single rotation rate can provide a fit nearly as good as that from a distribution of rotation rates.  The necessary rotation rate is very high, however, at $\Omega/\Omega_{\rm crit} \sim 0.7$, which is incompatible with the observed distribution of young stars \citep{Zorec+Royer_2012}.  This results from the fact that the slower rotators have already evolved onto the giant branch, leaving only the tail of the distribution on the main sequence turnoff \citep{Georgy+Granda+Ekstrom+etal_2014}.  Adopting an isochrone with a single rotation rate would have almost no effect on our inferred age, and would modestly degrade the agreement of the age distributions of the individual stars.  

The rotating model provides an excellent fit to the observed stellar density in both clusters, confirming their common age and composition.  The fit is qualitatively better than for either of the nonrotating isochrones, and accounts for the width of both the upper end of the main sequence and of the main sequence turnoff.  While the actual age inferred depends on the value of the convective overshooting parameter and on the exact composition of the stellar models, our results clearly favor an older age for both the Hyades and Praesepe than is currently accepted.  If we adopt an age of $\sim$800 Myr, the apparent intracluster age dispersion problem vanishes.  

\section{Conclusions} \label{sec:conclusions}

In this paper we have used a Bayesian color-magnitude dating technique including rotating stellar models to the Praesepe and Hyades open clusters.  We have shown that the clusters are strongly inconsistent with a single age at their spectroscopic composition if we neglect stellar rotation.  Including stellar rotation, however, makes the discrepancy vanish.  The Hyades and Praesepe are fully consistent with a single episode of star formation and the observed distribution of stellar rotation rates.  

However, the age we derive for the Hyades and Praesepe, $\sim$800 Myr, is older than the consensus age of $\sim$600-650 Myr.  This arises largely from the increase in main sequence lifetime with rotation.  To a lesser extent, our use of models with the updated, less metal-rich Solar composition of \cite{Asplund+Grevesse+Sauval+etal_2009} also increases the age.  Eliminating the modest core convective overshooting in the \cite{Georgy+Ekstrom+Eggenberger+etal_2013} models could also bring our ages closer to the accepted values.  However, this would likely degrade the agreement of the models with observed color-magnitude diagrams, as the overshoot parameter was tuned to match observations \citep{Ekstrom+Georgy+Eggenberger+etal_2012}.  Asteroseismic observations also require some degree of overshooting \citep{Moravveji+Aerts+Papics+etal_2015}. 

In spite of these caveats, Figure \ref{fig:HR} shows that the rotating model reproduces the entire upper main sequence remarkably well.  We have simply adopted and interpolated the rotating models of \cite{Ekstrom+Georgy+Eggenberger+etal_2012,Georgy+Ekstrom+Eggenberger+etal_2013} and fit for an age.  We have adopted the spectroscopic metallicity and an empirically-motivated rotation distribution, adding no additional free parameters to the model.  Including rotation removes the need for a large spread in ages and suggests that the Hyades and Praesepe may be significantly older than is currently thought.  If the older age is correct, it requires a recalibration of secondary age indicators like activity and rotation.

\acknowledgements{The authors thank Geza Kovacs and Lynne Hillenbrand for suggesting that we take a closer look at the Hyades and Praesepe, and for helpful comments on the manuscript.  We also thank Scott Tremaine for suggestions that made the statistical discussion much clearer.  This work was performed in part under contract with the Jet Propulsion Laboratory (JPL) funded by NASA through the Sagan Fellowship Program executed by the NASA Exoplanet Science Institute. This research has made use of the SIMBAD database and the VizieR catalogue access tool, operated at CDS, Strasbourg, France. }

\bibliographystyle{apj_eprint}
\bibliography{refs}

\end{document}